\documentclass[11pt]{article}

\usepackage{times}
\usepackage{natbib}
\bibpunct{(}{)}{;}{a}{,}{,}

\usepackage{fullpage}
\usepackage{graphicx}
\usepackage{amsmath,amssymb,amsthm}
\usepackage{bm}
\usepackage{latexsym}

\usepackage[pdftex,colorlinks=true,linkcolor=blue,citecolor=blue,urlcolor=blue,bookmarks=false,pdfpagemode=None]{hyperref}

\usepackage{url}
\makeatletter
\def\url@leostyle{%
  \@ifundefined{selectfont}{\def\UrlFont{\sf}}{\def\UrlFont{~\it\small\ttfamily}}}
\makeatother
\usepackage{verbatim}
\usepackage{fancyhdr}
\usepackage{setspace}
\usepackage{paralist}
\usepackage{boxedminipage}

\newcommand{\bv}{\begin{array}}
\newcommand{\ev}{\end{array}}
\newcommand{\bit}{\begin{itemize}}
\newcommand{\eit}{\end{itemize}}
\newcommand{\ben}{\begin{enumerate}}
\newcommand{\een}{\end{enumerate}}
\newcommand{\beq}{\begin{equation}}
\newcommand{\eeq}{\end{equation}}
\newcommand{\bvq}{\begin{eqnarray}}
\newcommand{\evq}{\end{eqnarray}}

\newtheorem{lemma}{Lemma}

\def\abovestrut#1{\rule[0in]{0in}{#1}\ignorespaces}
\def\belowstrut#1{\rule[-#1]{0in}{#1}\ignorespaces}
\def\abovespace{\abovestrut{0.20in}}

\def\belowspace{\belowstrut{0.10in}}

\begin{document}
\thispagestyle{empty}
\section*{\centerline{\Large Mixed membership analysis of genome-wide expression data}\vspace{20pt}\newline 
          \it \normalsize 
          Edoardo M. Airoldi, Princeton University \hfill (eairoldi@princeton.edu)\newline
          Stephen E. Fienberg, Carnegie Mellon University \hfill (fienberg@stat.cmu.edu)\newline
          Eric P. Xing, Carnegie Mellon University \hfill (epxing@cs.cmu.edu)}

\begin{abstract}
  Learning latent ``expression themes'' that best express complex patterns in a sample is a central problem in data mining and scientific research. For example, in computational biology we seek a set of salient gene expression themes that explain a biological process, extracting them from a large pool of gene expression profiles. 
  In this paper, we introduce probabilistic models to learn such latent themes in an unsupervised fashion.
  Our models capture ``contagion'', i.e., dependence among multiple occurrences of the same feature, using a hierarchical Bayesian scheme. Contagion is a convenient analytical formalism to characterize semantic themes underlying observed feature patterns, such as ``biological context''. We present model variants tailored to different properties of biological data, and we outline a general variational inference scheme for approximate posterior inference.  
  We validate our methods on both simulated data and realistic high-throughput gene expression profiles via SAGE.  Our results show improved predictions of gene functions over existing methods based on stronger independence assumptions, and demonstrate feasibility of a promising hierarchical Bayesian formalism for soft clustering and latent aspects analysis.\newline

\noindent\textbf{Keywords:}  Contagion processes, Poisson distribution, negative-binomial distribution, hierarchical Bayesian models, mixed membership, variability allocation, approximate posterior inference, mean-field approximation, serial analysis of gene expression (SAGE) data. 
\end{abstract}

\section{Introduction}
\label{sec:introduction}

 As a consequence of the information glut society faces, a fundamental issue in data mining and scientific pattern discovery is that of finding a useful representation of complex systems, which is amenable to mathematical and statistical learning analyses.  A useful representation would summarize the plethora of feature patterns observed in a sample with a small set of typical ``feature expression themes'' that replicate the variability of the observations.
 For example, in computational biology, we may seek a set of salient gene expression patterns, not directly observable, that explain a biological process from a large pool of observed gene expression profiles;  in text analysis, we may seek the set of latent topics, i.e., typical word distributions, that best explain a collection of documents. The set of latent themes for feature can then be used for further analyses about the behavior of the whole system.

 The task of identifying latent themes is essentially a clustering problem, where we have little or no information about the properties of the themes/clusters we seek.  For any given number of latent themes\footnote{Such "themes" usually correspond to parametric formulations of the feature generation process \cite{Airo:Fien:Jout:Love:2006}.}, we seek to allocate observed feature expressions to possible underlying themes; or, in other words, we need to cluster objects (e.g., genes or documents) that are similar in terms of their observed feature expression profiles into coherent themes.
 Among existing approaches to this task, \cite{Cai:Huan:Blac:Liu:2004} introduce a variant of $K$-means algorithm that minimizes non-standard scoring functions, based on the Chi-square statistic, and the Poisson distribution of feature expression rates---see Section \ref{sec:discussion} for more details. 
 This approach, however, constrains all feature expression levels measured on the same object to follow the expression profile typical of a single theme. \cite{Prit:Step:Donn:2000} relax this assumption, and posit that feature expression levels measured on the same object (in their case, the occurrences of a defined set of genetic polymorphisms in an individual) are mixtures of the expression profiles typical of several themes, i.e., population-specific frequency of each polymorphism. This model was independently re-discovered in the machile learning community \cite{Mink:Laff:2002,Blei:Ng:Jord:2003} with the goal of learning topics from a collection of documents---we refer to this popular model as the ``independence model'' in the remaining of this paper.
 %
 %
 %
 Recently, there has been a flurry of research on soft clustering in the machine learning and computational biology communities \cite{Cohn:Hofm:2001,Rose:Prit:Webe:etal:2002,Xing:Jord:Karp:Russ:2003,Grif:Stey:2004,Bunt:Jaku:2004}. The proposed models, however, are often unrealistic and fall short of replicating the true marginal variability profiles of the observations. In particular, as discussed in Section \ref{sec:analytical} and Section \ref{sec:empirical}, existing models appear to be unsuitable for the biological application (i.e., SAGE analysis) we concern in this paper. 

 In this paper, we introduce a hierarchical Bayesian formalism to address these problems. Briefly, our models
\begin{enumerate} 
 \item learn latent feature expression themes from data in an unsupervised fashion,
 \item enable domain-specific information to be incorporated in the form of priors on the hyper-parameters at the top of the hierarchy;
 \item assume that each feature may be instantiated under various themes to different degrees (i.e., feature emission probabilities are mixtures).
\end{enumerate}

Furthermore, we introduce the notion of {\em contagion}, which refers to the existence of dependences among subsequent occurrences of the same feature, when modeling objects on the basis of a generative process. Contagion is a convenient analytical formalism to capture richer variability profiles than current models allow for, and it characterizes plausible semantic themes, such as ``biological context'', underlying the observed feature patterns. In this paper, we present an analysis based on the contagion process of gene expression profiles measured via the SAGE technology. We show a new approach to summarize samples of gene expression data into latent gene expression themes, and compare our approach with extant algorithms. It is worth pointing out that the models we present here apply to a wider array of problems, e.g., the summarization of a collection of scientific publications into latent word frequency profiles, typically referred to as topics in the machine learning community.
 

Here is the plan for the rest of paper. We introduce the biological problem, and motivate the contagion process in Section \ref{sec:problem}. Then we present several variants of our model tailored to different properties of gene expression data in Section \ref{sec:contagion}.  
A general variational inference scheme for approximate posterior inference is outlined in Section \ref{sec:inference}. And we validate our methods on both simulated data and realistic high-throughput mouse retinal gene expression profiles via SAGE in Section \ref{sec:experiments}.
 

\section{The Biological Problem}
\label{sec:problem}

Serial analysis of gene expression (SAGE)~\cite{Vesc:Zhan:Voge:Kinz:1995} is a technology that quantitatively measure the copy numbers of mRNA transcripts, simultaneously for a large number of gene in a biological sample, such as a cell population or a tissue.
 
 
 A SAGE experiment begins by sampling a total of $B$ transcripts at random from a biological sample under some specific condition (e.g., a cell cycle stage), and then use $N$ gene-specific tags to probe the existence of possible genes in each of the $B$ transcripts. Let $X_b=(X_{b1}, X_{b2},\ldots,X_{bN})^T, X_{bn} \in \{0,1\}, \sum_n X_{bn} =1$ be a {\em unit-base} indicator vector recording the probing results for transcript $b$ (i.e., $X_{bn}=1$ indicates that gene $n$ is present on transcript $b$). The number of mRNA copies of a gene $n$, denoted by $Y_n$, and the vector of copy counts for all genes (i.e., an expression profile), $Y=(Y_{1}, Y_{2},\ldots,Y_{N})^T$, can then be simply expressed as: 
\begin{equation}
  Y_n=\sum_{b=1}^B X_{bn}, \quad \quad Y=\sum_{b=1}^B X_{b}.
\end{equation}
 Note that $Y_{n}$'s are each binomial distributed, controlled by gene-specific parameters $p_{1:N}$ each captures the probability of occurrence of gene on a random transcript, and a common sample size parameter $B$.
When multiple cellular conditions are of interest, for example, stage sequences in a cell cycle, we can index an expression profile with its sample condition, e.g., $Y^{t}$, for measurements obtained at time $t$.

 The main random quantities of interest are: 
 the observed \textit{gene expression levels} $Y_n^t$'s, for the $n$-$th$ gene at the $t$-$th$ epoch; 
 the observed \textit{gene expression profiles} $Y_n^{1:T}$'s, for the $n$-$th$ gene;
 and the latent \textit{gene expression themes}, e.g., $p_k^{1:T}$ or $\lambda_k^{1:T}$, for the $k$-$th$ theme, as defined in \cite{Prit:Step:Donn:2000} and in the basic model of Sections \ref{sec:poisson}, respectively.
 Technically, the latent gene expression themes are multivariate emission probabilities for the gene expression levels, conditionally on the ``active'' theme.  The notation we adopt puts forward the set of parameters underlying a specific distribution, e.g., $\lambda_k^{1:T}$ is a vector of Poisson rates, which control the expression levels of those genes that are expressed according to the $k$-$th$ theme.  For example, whenever the $n$-$th$ gene is expressed according to the $k$-$th$ theme we have
\[
 Y_n^{1:T} \sim \bigm[ Pois(\lambda_k^1), \dots, Pois(\lambda_k^T) \bigm].
\]

\subsection{Analytical Justifications of Contagion}
\label{sec:analytical}

 In the biological problem above, we often face situations where occurrences of the same gene under single and multiple conditions are not independent of one another, because they are sampled from a cell population or a tissue that provides a specific ``biological context''. Contagion processes provide a useful analytical mechanism to capture this notion.
 The two generative models we propose for analyzing temporal gene expression data $\{Y_n^{1:T}\}_{n=1}^N$, which instantiate the contagion process, are based on the a Poisson and a negative-binomial distribution of integer counts\footnote{For a review of various parameterizations, and the corresponding estimators we refer to \cite{Airo:Cohe:Fien:2005}, \cite{John:Kotz:Kemp:1992} and \cite{Kada:Shmu:Mink:etal:2006}.}, at different levels.
  
 Our choices were motivated by few main considerations.
 The Poisson distribution offers a computational advantage over the binomial distribution.  We can reasonably assume that the gene-specific probabilities of occurrence $p_{1:N}$ are very small, given that there is a large amount of transcripts present in a specific biological sample.  Consequently, it is reasonable to approximate the binomial probabilities with Poisson probabilities, as well as computationally efficient.
 The sampling algorithms underlying both the Poisson and negative-binomial distributions lead to marginal and conditional\footnote{Conditionally on the ``active'' theme.} distributions for the gene expression levels with desirable properties.  Assuming Poisson or negative-binomial conditional emission probabilities relaxes the assumption that, in the (sequential) sampling process described in Section \ref{sec:problem}, subsequent observed instances of the same gene tag are independent.  In fact, such independence leads to binomial conditional emission probabilities \cite{Prit:Step:Donn:2000}.
 The dependence among different observations of the same gene tag at the conditional level is a one characteristic of the notion of contagion we introduce.
 Another characteristic of our notion of contagion is found at the marginal level.  Recall that ideally we identify themes that can be interpreted as ``biological or functional contexts''.  Following the intuition that each gene may be expressed under multiple biological contexts to a different degree, we model the probability of observed gene expression levels, $Y_n^t$, as a mixture of conditional emission probabilities, where the gene-specific mixture weights given by the mixed membership vectors, $\theta_n$, are constant over time (or across experimental conditions).
 The mixing leads to marginal distributions that are more skewed than the corresponding conditional distributions.  This is the ``contagion effect'' more popular in the literature\footnote{Although this second characteristic of contagion processes is more common in the literature, there is an subtle point to notice in latent aspect models that feature independence of subsequent observed instances of the same gene tag \cite{Prit:Step:Donn:2000,Mink:Laff:2002,Blei:Ng:Jord:2003}.  Specifically, if we model themes as multinomial distributions, then Dirichlet distributed mixing weights will not alter the mean-to-variance ratio of the marginal distribution, which is still multinomial.  Rather, the main effect of mixing is an increased variability.} \cite{Simo:1955}.
  For example, in the case where the conditional probabilities are Poisson, their mixing would increase the variability of the expression levels.  A formal model of contagion that encodes this intuition is the negative-binomial model, which arises as an infinite Gamma mixture of Poisson distributions.
 These arguments support our distributional choices.  From a data analysis standpoint, the marginal distributions that encode contagion fit well the observed expression levels.
 
 To summarize, contagion processes are the result of latent regularities present in structured data, such as the SAGE profiles we study.  The inherent topical structure of the data, i.e., the fact that genes may be expressed under several latent themes, leads to hierarchical mixing of emission probabilities, and, ultimately, to the over-dispersion of gene expression levels.

\begin{table*}[t!]
\caption{Methods-of-Mometns estimates of negative-binomial parameters for gene expression levels in mouse retinal cells of at 10 different stages of development \cite{Cai:Huan:Blac:Liu:2004}.  A discussion of the estimators is given in \cite{Airo:Cohe:Fien:2005}.}
\label{tab:sage_stats}
\begin{center}
\begin{tabular}{crrrrr}
\hline
\abovespace\belowspace
 Epoch & mean~ & var.~~ & $\frac{\sqrt{\hbox{var.}}}{\hbox{mean}}$ & $\sigma$~~~~~~~~~~ & $\xi$~~~~~~~~~~ \\
\hline
\abovespace
  1 &  30.1172 & 150.8648 &  2.2381 & 11.1733 $\pm$ 0.3655 &  4.3000 $\pm$ 0.2155 \\
  2 &  26.5542 & 163.8892 &  2.4843 &  9.8514 $\pm$ 0.4075 &  6.1021 $\pm$ 0.3304 \\
  3 &  28.1718 & 155.4820 &  2.3493 & 10.4516 $\pm$ 0.2936 &  2.9376 $\pm$ 0.1448 \\
  4 &  31.5446 & 204.2503 &  2.5446 & 11.7029 $\pm$ 0.3267 &  3.2591 $\pm$ 0.1588 \\
  5 &  26.0307 &  94.4013 &  1.9043 &  9.6572 $\pm$ 0.4154 &  6.4720 $\pm$ 0.3562 \\
  6 &  26.6489 &  82.0171 &  1.7543 &  9.8866 $\pm$ 0.2118 &  1.5748 $\pm$ 0.0795 \\
  7 &  27.3122 &  82.0405 &  1.7331 & 10.1327 $\pm$ 0.2491 &  2.1565 $\pm$ 0.1066 \\
  8 &  25.1990 &  53.6102 &  1.4586 &  9.3487 $\pm$ 0.2637 &  2.6407 $\pm$ 0.1319 \\
  9 &  27.1513 &  89.7169 &  1.8178 & 10.0730 $\pm$ 0.4472 &  7.2014 $\pm$ 0.4008 \\
\belowspace
10 &  20.8160 &  81.2509 &  1.9757 &  7.7226 $\pm$ 0.5975 & 16.8959 $\pm$ 1.3156 \\
\hline
\end{tabular}
\end{center}
\end{table*}

\subsection{Empirical Evidence}
\label{sec:empirical}

Our motivating example is the set of mouse retinal SAGE libraries analyzed in \cite{Cai:Huan:Blac:Liu:2004}. The raw mouse retinal data consists of 10 SAGE libraries (38,818 unique genes that appeared more than twice in the sample) from developing retina taken at 2-day intervals, ranging from embryonic day to postnatal day, and adult, for total of 10 epochs \cite{Blac:Harp:Trim:Cai:2004}. Of the 38,818 genes, 1,467 that appeared more than 20 times in at least one of the 10 libraries were selected.  These 1,467 genes were purported as the potentially most biologically relevant because of their high frequency of occurrence. The data analyzed in this paper consists of the pool of observed expression profiles $(Y_{n}^{1}, Y_{n}^{2}, \dots, Y_{n}^{10})$ for the 1,467 selected genes, measured at ten epochs during the development period.

We tested the distributional intuitions we discussed in \S\ref{sec:analytical} on the SAGE data.  In Table \ref{tab:sage_stats} we report summary statistics and estimates for the negative-binomial parameters described in \cite{Airo:Cohe:Fien:2005}.
Our exploratory data analysis confirms the expected over-dispersion of the gene counts, suggested by the ``mixture of Poisson distributions'' hypothesis.  Moreover, the estimates of the extra-Poissonness parameter $\delta$ are all positive\footnote{Recall that as $\delta \rightarrow 0$ the negative-binomial density degenerates into a Poisson density.} with very high probability, as indicated by a quick inspection of the corresponding standard deviations.
Lastly, note that the log transformation $\zeta = \log(1+\delta)$ is effective in reducing the heavy tail of the distribution of $\delta$.  Thus, we prefer to work on the $\zeta$ scale, where positing a simple prior is sensible.
 
In conclusion, the SAGE data we analyzed are not under-dispersed (i.e., variance $<$ mean), as is implied by treating the random variables $\{X_n^{1:B}\}$ 
as Bernoulli processes \cite{Prit:Step:Donn:2000,Rose:Prit:Webe:etal:2002}.  Such an assumption leads to clustering models based on Multinomial latent profiles and binomial emission probabilities for feature counts \cite{Blei:Ng:Jord:2003,Grif:Stey:2004,Bunt:Jaku:2004}, which are not often warranted.

\section{Contagion Processes}
\label{sec:contagion}

 In this section we introduce two hierarchical Bayesian generative processes for clustering SAGE data into expression profiles in an unsupervised fashion.  These models capture ``biological context'' through the notion of contagion. 
 Recall that we observe sequences of gene tag counts $(Y_{n}^{1}, Y_{n}^{2}, \dots, Y_{n}^{T})$ that measure the expression level (i.e., the abundance) of the $n$-$th$ gene in the target cell or tissue across epochs 1 though $T$.
 In the models below we assume there is a fixed number, $K$, of latent expression profiles, and that genes are expressed under different profiles to different degrees.

\begin{figure*}[t!]
\begin{center}
  \centering
   \includegraphics[width=15cm]{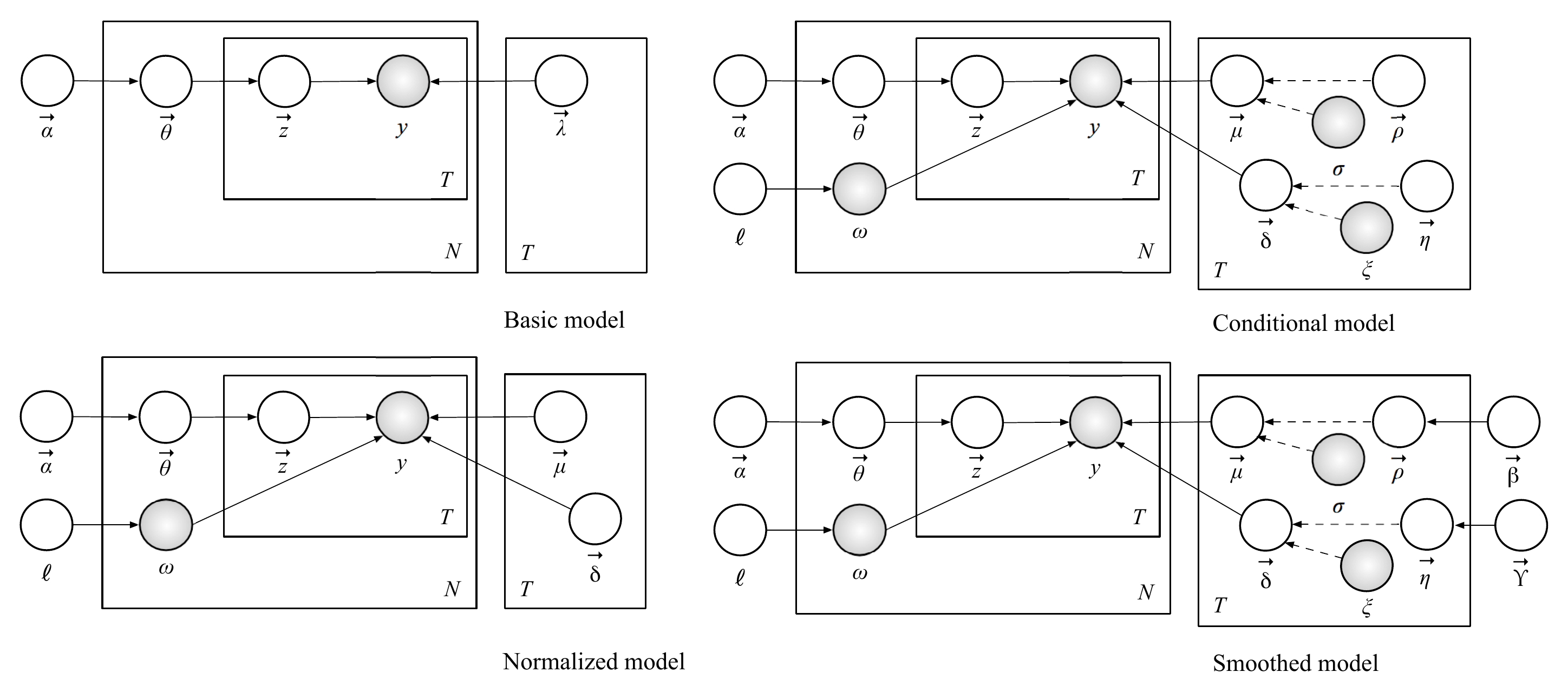}
\caption{Graphical representation of the generative processes of contagion based on the Poisson (top left) and negative-binomial sampling schemes.  The representation for the processes of contagion based on the Poisson sampling scheme for the non-basic models are easily obtained, by removing the part of the graphical models depending on $\delta$.  In fact, recall that $\delta$ is the extra-Poissonness parameter, and as $\delta \rightarrow 0$ the negative-binomial density converges to the corresponding Poisson limit.  We refer to \cite{John:Kotz:Kemp:1992} for more details.}
\label{fig:dinbmodels}
\end{center}
\end{figure*} 

\subsection{Poisson Generative Process}
\label{sec:poisson}

The first generative process we propose is based on the Dirichlet and Poisson distributions. There are four flavors of the Dirichlet-Poisson generative process: basic (bDiP), normalized (nDiP), conditional (cDiP), and smoothed (sDiP).
 In the ``basic'' model we explicitly posit the ``mixed-membership'' of genes to latent profiles by assigning to each gene a Dirichlet vector of probabilities, $\bm{\theta}_{n}$.
 In order to generate the observed expression levels $Y_n^{1:T}$ of the $n$-$th$ gene, assuming $K$ latent expression profiles, we proceed as follows.
\begin{itemize}
  \item[1.] Sample $\bm{\theta}_{n} \sim Dirichlet_{\,K} \, (\bm{\alpha})$ 
  \item[2.] For each epoch $t=1,\dots,T$
   \begin{itemize}
    \item[2.1.] Sample $\bm{z}_{n}^{t} \sim Multinomial \, (\bm{\theta}_{n},1)$
    \item[2.2.] Sample $y_{n}^{t} \sim Poisson \, (\lambda_{tk} | z_{nk}^{t}=1)$.\\
   \end{itemize}
\end{itemize}

 The genes are the sampling units in SAGE experiments, and the total volume of their expressions often vary over time. We want to recover "calibrated" expression profiles that do not depend on the total expression volume.  Therefore, we posit the ``normalized'' model in order to rescale the samples (i.e., the genes) according to their different sizes (the total expression volumes), and ultimately improve the parameter estimates.
In the basic model, the matrix $\bm{\lambda}\equiv\{\lambda_{tk}\}$ contains the rates that govern the expression level of genes at $T$ different epochs for each of the $K$ different latent profiles.  In the normalized model, the expected expression level of a gene $\tau_n$, at time $t$ for profile $k$, is
\beq
 \lambda_{tk} = \omega_{n} \cdot \mu_{tk},
\eeq
 where $\omega_{n}$ is scalar and observed, and denotes the total expression level of gene $\tau_{n}$ as a multiple of a fixed total expression level $\beta$ used as a reference expression level.  This new parameter $\beta$ may a fixed pre-determined value,   estimated via, e.g., empirical Bayes \cite{Carl:Loui:2005}, or given a distribution as part of a full Bayesian analysis \cite{Airo:Ande:Fien:Skin:2006}. 

In both the basic and the normalized models above, the rows of the parameter matrices $\bm{\lambda}$ and $\bm{\mu}$ control the rates at which genes are expressed.  In particular, $\lambda_{tk}$ and $\mu_{tk}$ encode the expected expression level of genes at time $t$ for profile $k$.  Since profiles are by definition not observable, none of these parameters can be estimated directly from the data.
  
 We reparameterize the rows of the normalized rate matrix $\bm{\mu}$ with the sum/ratio parameterization, i.e., for every epoch $t$ we transform 
\beq
\label{eq:sum_ratio}
  (\mu_{t1}, \mu_{t2}, \dots, \mu_{tK}) \longrightarrow (\sigma_{t}, \rho_{t1}, \rho_{t2}, \dots, \rho_{tK}),
\eeq
 where the sum parameter $\sigma_{t} := \sum_{k=1}^K \mu_{tk}$, the ratio parameters $\rho_{tk} := \frac{\mu_{tk}}{\sigma_{t}}$, and the constraint that $\sum_{k=1}^{K} \rho_{tk} = 1$ makes the ratio parameter $\rho_{tK}$ redundant for each $t$.
 
 This reparameterization leads to the ``conditional'' model, where the sum parameters $(\sigma_{1}, \sigma_{2}, \dots, \sigma_{T})$ are directly estimable from the data, and we can carry out inference conditionally on them.  This is possible since the parameters $\sigma_{t}$ encode the total normalized  expression level at time $t$, sum of the expression levels over the profiles, which is an observable quantity as it does not depend on the latent profiles.  Conditioning on the MLEs for the total expression parameters, $\sigma_{t}$, leads to a new allocation problem where we need to infer the differential expression levels of genes under the $K$ profiles.  In other words, we need to ``split'' the total expression level at each time $t$, given by a  direct estimate of $\sigma_{t}$, among the latent profiles.

 Last, we introduce the ``smoothed'' model, where we posit a posit a prior for  the differential expression rate parameters to smooth the estimates.
It is possible to posit a prior distribution on the total expression rate parameters as well, but we choose not to.  A brief analysis of the observed total rates suggests it is appropriate to apply a logarithmic transformation on them to stabilize the variability, and one can introduce a Gaussian prior on the transformed rates; however, an inspection of the total rates $\sigma_t$ over time (see Table \ref{tab:sage_stats}) suggests that some other phenomenon is possibly going on, which leads to a decreasing occurrence of the genes in the SAGE libraries.  Therefore we choose to use the observed total rates to inform our inferences directly, as in the conditional model\footnote{Smoothing the overall rates $\{\sigma_{tk}\}$ would impose a model on data that we would not be able to justify, since we do not have an intuition of why the overall rates are declining.  This would cast some doubts on the interpretability of the inferences such a model would lead to.}.
However, in the smoothed model we sample the differential expression levels
\[
 \bm{\rho}_{t~\cdot} \sim Dirichlet_{\,K} \, (\bm{\beta})
\]
for each epoch $t=1,2,\dots,T$.  See Figure \ref{fig:dinbmodels}.
 
 In conclusion, the Dirichlet-Poisson generative process possesses a few advantages: (1) this sampling scheme encodes contagion in the sense that multiple occurrences of the same gene tag at the same epoch depend on one another, under a specific latent expression theme;  (2) this sampling scheme arises naturally in the biological experiments we are interested in as we discussed in \S\ref{sec:analytical};  (3) computing Poisson probabilities is computationally more efficient than computing binomial probabilities, since we do not have to evaluate binomial coefficients.

\subsection{Negative-Binomial Generative Process}

The generative process of contagion based on the negative-binomial sampling scheme is similar in spirit to the previous one based on the Poisson sampling scheme.  A formal treatment of models along this line, however, would involve tedious parameterization details that is beyond the scope of this paper.
Intuitively, the negative-binomial distribution has two parameters that control mean and variance, and the variance is greater than the mean---this is a useful feature to capture the observed over-dispersion of gene expression levels.  Its density can be written as a Poisson density with an extra parameter $\delta$ that controls the amount of extra-Poisson variability.  Such a version of the density is our starting point,
\[
 NB \bigm( y_n^t \bigm| \omega_{n}\mu_{t}, \omega_n\delta_t) = \frac{\Gamma(y_n^t+\kappa_t)}{y_n^t! \Gamma(\kappa_t)} \, \frac{(\omega_{n} \delta_t)^{y_n^t}}{(1+ \omega_{n} \delta_t)^{(y_n^t+\kappa_t)}},
\]
where $\kappa_t := \frac{\mu_t}{\delta_t}$ for convenience of notation.
 In normalized model, $\{\mu_{tk}\}$ are the profile-specific Poisson rates and $\{\delta_{tk}\}$ are profile-specific extra-Poissonness parameters.  We then introduce the conditional model, where we apply the sum/ratio parameterization of equation \ref{eq:sum_ratio} to both sets of parameters to obtain mappings
\bvq
 (\mu_{t1}, \mu_{t2}, \dots, \mu_{tK}) &\longrightarrow& (\sigma_{t}, \rho_{t1}, \rho_{t2}, \dots, \rho_{tK}) \\
 (\delta_{t1}, \delta_{t2}, \dots, \delta_{tK}) &\longrightarrow& (\xi_{t}, \eta_{t1}, \eta_{t2}, \dots, \eta_{tK}).
\evq
 Finally, in the smoothed model we sample the differential extra-Poissonness parameters
\[
 \bm{\eta}_{t~\cdot} \sim Dirichlet_{\,K} \, (\bm{\gamma})
\]
for each epoch $t=1,2,\dots,T$.  See Figure \ref{fig:dinbmodels}.

\section{Approximate Posterior Inference}
\label{sec:inference}
 
 Inference in these models is a challenging task.  In fact, in order to obtain the posterior for the latent variables,
\begin{equation}
 \label{eq:post}
 p \bigm( \{ \theta_{n}, z_{n}^{1:T} \}_{n=1}^N \bigm| \{ y_n^{1:T} \}_{n=1}^N, \alpha, \{ \lambda_{k}^{1:T} \}_{k=1}^K \bigm) = \frac{p \bigm( \{ \theta_{n}, z_{n}^{1:T} \}_{n=1}^N, \{ y_n^{1:T} \}_{n=1}^N \bigm| \alpha, \{ \lambda_{k}^{1:T} \}_{k=1}^K \bigm)}{p \bigm( \{ y_n^{1:T} \}_{n=1}^N \bigm| \alpha, \{ \lambda_{k}^{1:T} \}_{k=1}^K \bigm)},
\end{equation}
 we need to compute the likelihood of the data, which is given by an integral with no closed form solution.  The quantity at the denominator on the right hand side of Equation \ref{eq:post}, for example, is the likelihood of the data we need in order to solve the Bayes problem in the basic model of Section \ref{sec:poisson}.

 In order to perform inference in these models, we develop a mean-field approximation to the posterior that involves the substitution of an integrable lower bound for the likelihood.
 The mean-field approximation involves positing a simple distribution, $q$, over the latent variables, which depends upon an extra set of (variational) free parameters, $\{\nu_{n}, \phi_{n}^{1:T}\}_{n=1}^N$.  The free parameters are then set to minimize the Kullback-Leibler divergence between the true and approximate posteriors.
 This is equivalent to maximizing a lower bound for the likelihood within each E-step, over the free parameters, and then compute pseudo-exprectations for the latent variables off the tight lower bound.
 The overall inference algorithm is a variational EM scheme, which employs the mean-field approximation to carry out the E-step, as discussed above, and alternates with a regular M-step, where the pseudo maximum likelihood estimates of the model parameters, e.g., ($\alpha, \{ \lambda_{k}^{1:T} \}_{k=1}^K$) for the basic model, are revised by further maximizing the lower bound for the likelihood.  We iterate these two steps till convergence.

 The variational EM scheme just described practically translates into a coordinate ascent algorithm, where parameters are naturally organized into batches with similar semantics.  The parameter updates corresponding to the model variants we considered here are summarized in Table \ref{tab:inference}.
 
\subsection{A General Bayesian Formalism for Latent Aspects Analysis}

 The variational inference scheme we developed is quite general.  In fact, the free parameter updates (that are used to maximize the lower bound for the likelihood within each E-step) take a generic form applicable for all different conditional emission probability function we concern, e.g., Table \ref{tab:inference}.
 Furthermore, for a generic conditional emission probabilities $p ( y_n^t | \beta_k^t )$ for all $(n,t,k)$, with parameter set $\{ \beta_k^{1:T} \}_{k=1}^K$, we obtain the generic the free parameter updates
\[
 \phi_{ntk}^* \propto \Upsilon \cdot p \bigm( y_n^t \bigm| \beta_k^t \bigm),
\]
 where $\Upsilon := e^{\mathbb{E}_q [\log \theta_{nk}]}$ as in Table \ref{tab:inference}.  The updates for $\nu^*_{nk}$ do not change.

 The generality of the approximate E-step in latent aspects analysis that feature one latent group indicator, $z_n^t$, for each gene-epoch pair $(n,t)$ is due the specific hierarchical formulation of our models.  Such a formulation posits exchangeable measurements on features, e.g., gene expression levels at each epoch.  Different conditional emission probabilities only lead to different estimators for the corresponding parameters, $\{ \beta_k^{1:T} \}_{k=1}^K$, in the M-step.
 
\begin{table*}[t!]
\caption{In the table below we summarize the parsimonious mean-field approximation for the various models.  The parsimonious mean-field approximation posits one latent expression profile indicator $z$ for each (gene,epoch) pair.  Note that $\Upsilon := e^{\mathbb{E}_q [\log \theta_{nk}]}$, and $Po$, $NB$, are short for $Poisson$, and $Negative$-$Binomial$, respectvely.  $^{**}$ Alternatively use the Method of Moments described in \cite{Airo:Cohe:Fien:2005} pretending to observe pseudo counts $\{ \phi_{nk}^t \cdot y_{n}^t \}$ as the expression levels of the $n$-$th$ gene according to the $k$-$th$ latent theme.}
\label{tab:inference}
\begin{center}
\begin{sc}
\begin{tabular}{lll}
\hline
\abovespace\belowspace
          & Poisson & Negative-Binomial \\
\hline
\abovespace
 Basic & $\nu_{nk}^*=\alpha_k + \sum_{t} \phi_{ntk}$ & \\
       & $\phi_{ntk}^* \propto \Upsilon \cdot Po\bigm( y_n^t\bigm| \lambda_{tk} \bigm)$ & \\
       & $\lambda_{tk}^* = \frac{\sum_n \phi_{ntk} y_{n}^{t}}{\sum_n \phi_{ntk}}$ & \\
\belowspace
       & $\alpha_{k}^*$ with Newton-Raphson & \\ \hline
\abovespace
 Norm. & $\nu_{nk}^*=\alpha_k + \sum_{t} \phi_{ntk}$ & $\nu_{nk}^*=\alpha_k + \sum_{t} \phi_{ntk}$ \\
       & $\phi_{ntk}^* \propto \Upsilon \cdot Po\bigm( y_n^t\bigm| \omega_n\mu_{tk} \bigm)$ & $\phi_{ntk}^* \propto \Upsilon \cdot NB\bigm( y_n^t\bigm| \omega_n\mu_{tk} \bigm)$ \\
       & $\mu_{tk}^* = \frac{\sum_n \phi_{ntk} y_{n}^{t}}{\sum_n \phi_{ntk} \omega_n}$ & $\mu_{tk}^* = \frac{\sum_n \phi_{ntk} y_{n}^{t}}{\sum_n \phi_{ntk} \omega_n}$ \\
       & & $\delta_{tk}^* = L$-$BFGS^{~**}$ \\
\belowspace
       & $\alpha_{k}^*$ with Newton-Raphson & $\alpha_{k}^*$ with Newton-Raphson \\ \hline
\abovespace
 Cond. & $\nu_{nk}^*=\alpha_k + \sum_{t} \phi_{ntk}$ & $\nu_{nk}^*=\alpha_k + \sum_{t} \phi_{ntk}$ \\
       & $\phi_{ntk}^* \propto \Upsilon \cdot Po\bigm( y_n^t\bigm| \omega_n\sigma_t\rho_{tk} \bigm)$ & $\phi_{ntk}^* \propto \Upsilon \cdot NB\bigm( y_n^t\bigm| \omega_n\sigma_t\rho_{tk} \bigm)$ \\ 
       & $\rho_{tk}^* = \frac{\sum_n \phi_{ntk} y_{n}^{t}}{\sum_n \phi_{ntk} \omega_n \sigma_t}$ & $\rho_{tk}^* = \frac{\sum_n \phi_{ntk} y_{n}^{t}}{\sum_n \phi_{ntk} \omega_n \sigma_t}$ \\
       & & $\eta_{tk}^* = L$-$BFGS^{~**}$ \\
\belowspace
       & $\alpha_{k}^*$ with Newton-Raphson & $\alpha_{k}^*$ with Newton-Raphson \\ 
\hline
\end{tabular}
\end{sc}
\end{center}
\end{table*}

\section{Experiments}
\label{sec:experiments}

 A non-trivial difference in the generative process with respect to the ``independence model'' in \cite{Prit:Step:Donn:2000,Mink:Laff:2002,Blei:Ng:Jord:2003} has far reaching implications for applications.  For example, models of contagion provide a better fit for data with realistic mean-to-variance marginal ratios, such as that in biological applications to SAGE.  A better fit helps recovering more precise mixed memberships of genes to themes, as well as finding tighter clusters, with respect to the independence model.

 In this section, we support our claims with 3 sets of experiments:
 (1) In simulative experiments, we showed that DiP is better at recovering membership than the independence model when realistic SAGE mean/variance ratio holds; 
 (2) in small samples bearing realistic SAGE characteristics, although the recovered clusters differ only slightly, the estimated mixed-membership are sharper using DiP than with the independence model (the PoissonL in \cite{Cai:Huan:Blac:Liu:2004} does not facilitate estimation);  
 (3) in a real dataset, we recovered meaningful gene expression profiles according to an empirical evaluation scheme adopted in \cite{Cai:Huan:Blac:Liu:2004} (e.g., rhodospin and photoreceptors in same cluster), and obtained a reasonable estimate of the total number of salient expression themes. As in many biological clustering task, objective comparison of clustering results are difficult, but the fact that DiP gets less fragmented clustering (15 versus 30 clusters by PoissonL), and sharper cluster mixed-membership estimates suggests that it is a more reliable theme identification model. 
 
\subsection{Simulated Data}

 We first validate our models by examining to what extend they can recover the mixed-membership probabilities $\{\theta_{n}\}$, i.e., the soft cluster assignments of each gene, under various simulated conditions. We generated the ground truth using our generative processes, and we focused on scenarios where the ``mean'' expression level at the various epochs was lower than its corresponding ``variance''--- a realistic biological experimental scenario. 
We compare our models, normalized DiP and conditional DiP, with two other methods, the independence model \cite{Prit:Step:Donn:2000,Mink:Laff:2002,Blei:Ng:Jord:2003}, and the PoissonL model \cite{Cai:Huan:Blac:Liu:2004}.
 Our models yield higher likelihoods of expression profiles in the test set (not shown), and more accurate predictions of the latent theme id of each gene based on their observed expression levels.
 Out of 1000 genes we simulated, for example, nDiP and cDiP achieved 75.95\% and 70.32\% accuracy, respectively, whereas the independence model reached only 63.25\%.
 %

\begin{figure}[b!]
\begin{center}
  \centering
   \includegraphics[width=15cm]{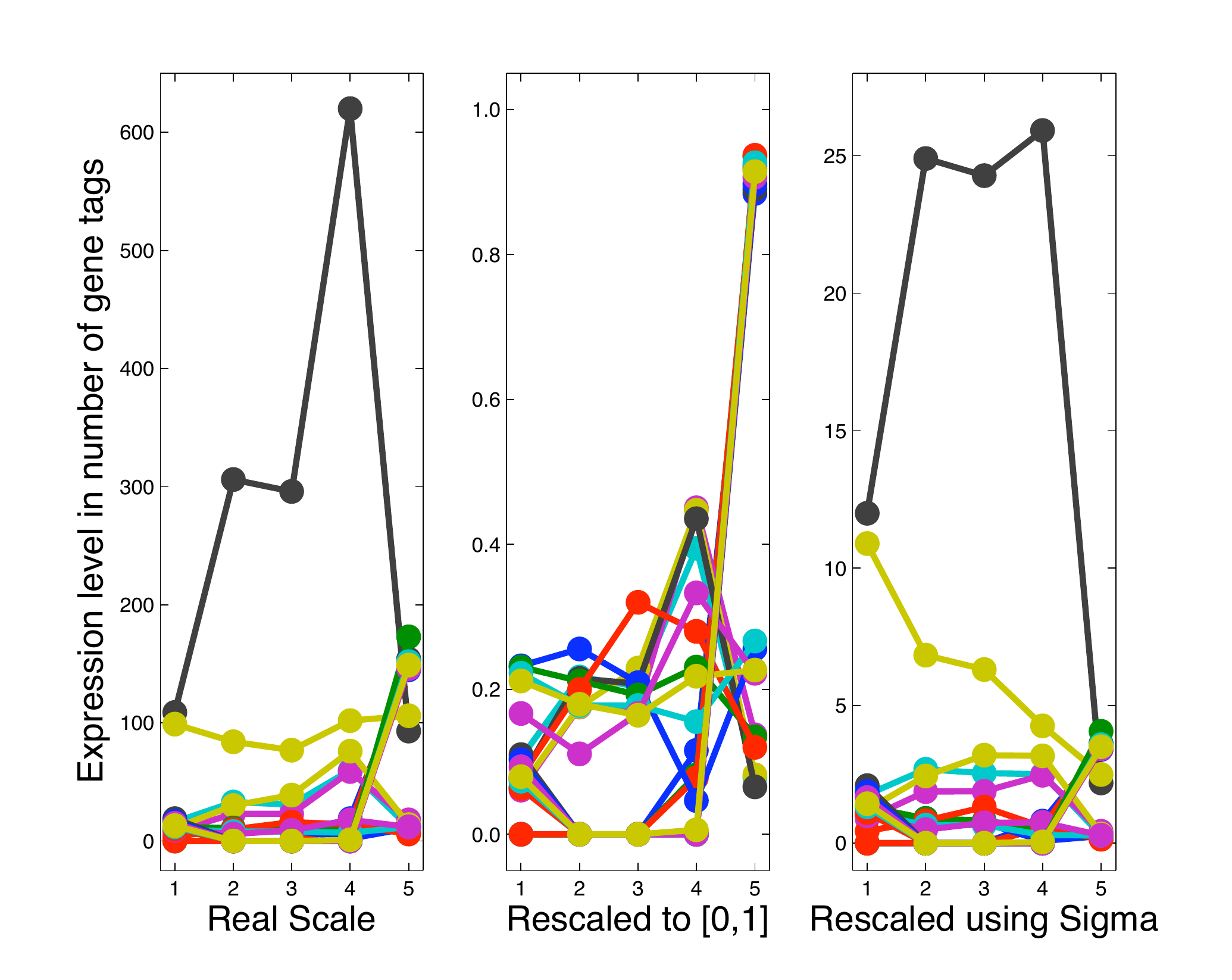}
\caption{The raw example data in \cite{Cai:Huan:Blac:Liu:2004}, on the original expression scale (left); on a normalized expression scale, by gene, into $[0,1]$ (center); and on a normalized expression scale, by epoch, using $\hat \sigma_{1:T}$ (right).}
\label{fig:ex_chw_data}
\end{center}
\end{figure} 

\subsection{A 20-gene Synthetic Data Set}

 Here we report our analysis of a small dataset used in \cite{Cai:Huan:Blac:Liu:2004}, which contains the expression profiles of 20 genes over 5 temporal epochs. Eighteen of the 20 genes belong to one of 4 clusters (temporal themes), and the 2 remaining two are identified as outliers. 

\begin{figure}[b!]
\begin{center}
  \centering
   \includegraphics[width=15cm]{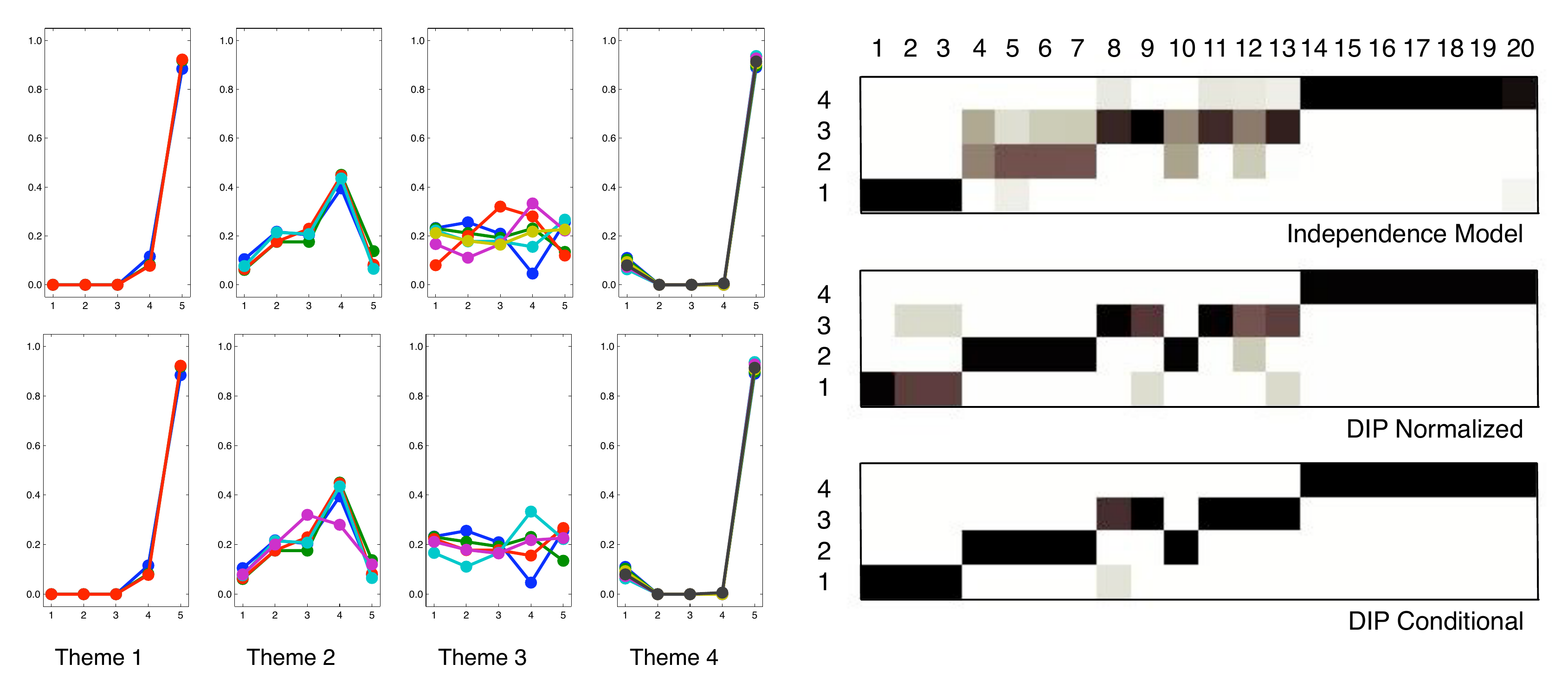}
\caption{Left: Latent gene expression themes learned by different algorithms. Top: 4 themes (numbered 1 to 4 from left to right) learned by PoissonL and the independence model. Each theme is represented by the expression profiles of all the genes assigned to that theme base on MAP prediction using the estimated mix-membership vector $\theta_n$. In this case, PoissonL and the independence model give the same membership prediction. Bottom: The 4 themes discovered by normalized DiP and conditional DiP. Note that due to overlap of the profile curves, the "occupancy" number of each theme is not apparent here. But in Fig.~\ref{fig:ex_chw_thetas}, one can see it more clearly.  Right: The estimated membership probabilities, $\{\hat \theta_{nk}\}$, for the independence model (top), nDiP (middle), and cDiP (bottom). Each row correspond to a theme, and each column corresponds to a gene. The color shades of the cells correspond to values ranging from 1 (black) to 0 (white).  The panel shows that cDiP yields the sharpest estimates.}
\label{fig:ex_chw_thetas}
\end{center}
\end{figure}

 The expression profiles are generated from 6 different latent themes, or clusters, which the authors reduce to 4 by ignoring the abundance of the gene tags observed on the transcripts sampled at each epoch.  In particular, there are 3 profiles from theme 1, 4 from theme 2, 6 from theme 3, and 6 from theme 4.  The raw data is plotted in Figure \ref{fig:ex_chw_data} on various scales.
 Among the profiles from theme 2, there is 1 with 10 times as many gene tags as the others, and similarly for theme 3---number 7 and number 13 in Figure \ref{fig:ex_chw_thetas}.  Note that these 2 profiles are ``more expressed'' but they follow an expression theme similar to the other expression profiles in the respective clusters.

 In Figure \ref{fig:ex_chw_thetas}, we display the 4 themes learned by the normalized and conditional DiP models (bottom-left panel), versus those learned by PoissonL \cite{Cai:Huan:Blac:Liu:2004} and the independence model (top-left panel). A rough eyeballing shows that the gene expression themes learned by DiPs and the two competing methods are similar. However, a close examination reveals the following. Arguably, we obtain a more compact themes 3, as revealed by the lower degree of dispersion among genes assigned to this theme; but for theme 2, the genes assigned to it by the independence model and PoissonL are slightly more consistent. Overall, the software clustering assignment of each gene are compatible across all 4 algorithms, and as shown in Figure \ref{fig:ex_chw_thetas}), but the mixed-membership probabilities inferred by the DiPs for each gene are sharper. If we compare the MAP assignment of each gene to a single most probable themes, the 19 of the 20 genes are consistent across all 4 algorithms, and their assignments agree with the true themes label given by the original dataset. The remain one, gene no. 10, is intriguing. It has an expression profile, $\{Y_{10}^{1:5}\} = (4, 10,16, 14, 6)$, and is originally labeled as from theme 2, $\{\lambda_{2}^{1:5}\} = (10, 30, 30, 60, 10)$. Apparently profile $\{Y_{10}^{1:5}\}$ exhibits great variability with respect to its supposedly underlying theme. Using DiP, we infer the label of gene no. 10 to be theme 3, which has a prototype profile $\{\lambda_{3}^{1:5}\} = (10, 10, 10, 10, 10)$, and indeed we found much of the variability in gene 10 is related to the overall abundance of all genes in different epochs, rather then its intrinsic trend. So we feel this assignment is arguable more plausible the the purported theme 2. As shown in Figure \ref{fig:ex_chw_thetas}), the independence model inferred a split assigned, about equally probable to theme 2 and 3. 

 To summarize, this little example is meant to show the role of realistic model properties in latent allocation tasks. The intuition is that if the model cannot express, on average, the salient properties of the data, then it may lead to artifactual effects.  Specifically, the unexplained variability will need to find a ``place-holder'', and it will typically tend to increase the variability of parameter estimates.

\subsection{Mouse Retinal SAGE Profiles}

\begin{figure}[b!]
\begin{center}
  \centering
   \includegraphics[width=15cm]{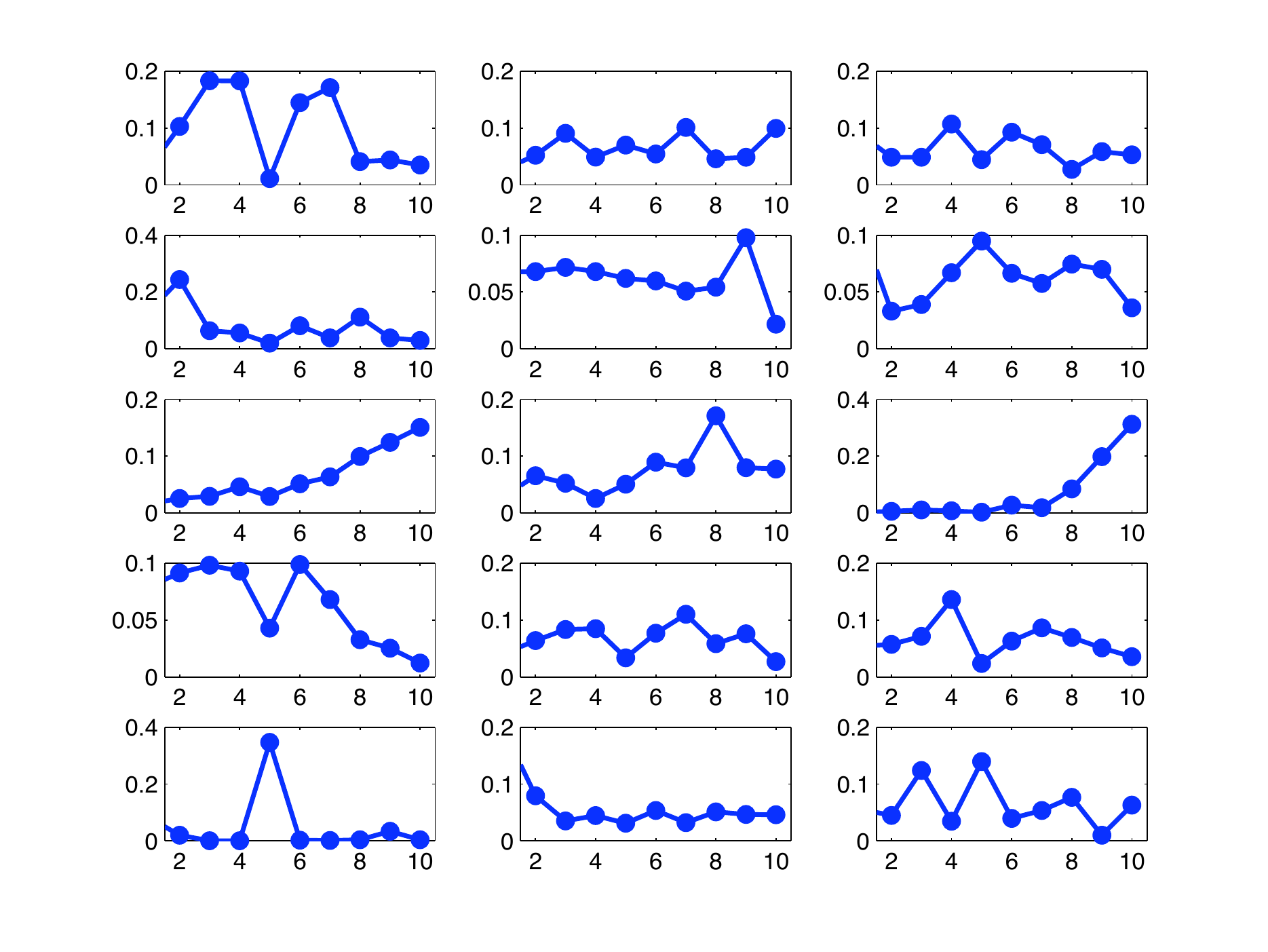}
\caption{Gene expression themes learned from mouse retinal SAGE using conditional DiP.}
\label{fig:sage_dip_mouse}
\end{center}
\end{figure} 

 Here we go back to the motivating case study we introduced in Section \ref{sec:empirical} --- the mouse retinal SAGE libraries analyzed in \cite{Cai:Huan:Blac:Liu:2004}, which contains 38,818 unique genes for total of 10 epochs. We first perform model selection via a five-fold cross validation, to estimate the plausible number of latent themes that best explain the data. The held-out likelihood peaked at 15 themes for cDiP, and 10 for the independence model. Figure~\ref{fig:sage_dip_mouse} shows the prototype gene expression profile for each of the 15 themes due to cDIP. The variance of each theme are not shown, because in many cases they are so small that the variance-bars are masked by the "dot" symbol in our plots.  Notably,  we found that the magnitude of the held-out likelihood for cDiP is about ten times larger (on the log scale) than that for the independence model, suggesting better a fit of DiP to the data.  Furthermore, the corresponding mixed-membership estimates of $\{\theta_n\}$ are more sharply peaked (as seem before in Figure~\ref{fig:ex_chw_thetas}).  This is also confirmed by the estimates of Dirichlet hyper-parameter, $\hat \alpha_{Indep} = 1.355$ versus $\hat \alpha_{DiP} = 0.066$.  
 The themes (or clusters) shown in Figure \ref{fig:sage_dip_mouse} indeed lead to reasonable predictions of mouse retinal gene functions.  For example, a preliminary biological validation of our clustering based on the GO annotation shows correlation between the latent themes and the functions for genes such as photoreceptors and rhodospin, i.e., genes with similar functional annotations tend to fall into the same theme in our analysis. An in-depth analysis of the biological significance of these clusters is given elsewhere.
 

\section{Related Work}
\label{sec:discussion}

 Here we discuss the connection between our algorithm and the PoissonC and PoissonL algorithms introduced by \cite{Cai:Huan:Blac:Liu:2004}.
 In the problem at hand we want to allocate the observed temporal expression profiles $\{Y_{n}^{1:T}\}_{n=1}^N$ into, say, $K$ themes or clusters.
 Recall that the $K$-means unsupervised clustering algorithm searches for $K$ mean profiles $m_{1:K}$ that minimizes
\[
 \label{eq:kmeans}
 MSE = \frac{1}{N} ~ \sum_{k=1}^K \sum_{n=1}^N ~ \mathbb{I} \bigm(y_{n}^{1:T} \in k \bigm) \left\| y_{n}^{1:T} - m_k \right\|^2.
\]
 That is, the mean profiles $m_{1:K}$ are centers of respective clusters in the sense of Euclidean norm.
 The PoissonC and PoissonL algorithms introduced by \cite{Cai:Huan:Blac:Liu:2004} substitute the euclidean norm in the equation with the chi-squared score,
\bvq
 \chi^2 (n,k) = \sum_{t=1}^T ~ \frac{\bigm( y_{n}^t - \hat \mu_{tk} ~ \hat \omega_{n} \bigm)^2}{\hat \mu_{tk} ~ \hat \omega_{n}},
\evq
 and the negative log-likelihood,
\bvq
 \ell (n,k) = - \sum_{t=1}^T ~ \log \left( \frac{e^{-(\hat \mu_{tk} ~ \hat \omega_{n})} ~ (\hat \mu_{tk} ~ \hat \omega_{n})^{y_{n}^t}}{y_{n}^t!} \right),
\evq
 respectively.
 Our normalized model based on the Poisson distribution is an extension of the PoissonL algorithm, where we introduce Dirichlet distributed mixed-membership vectors, $\theta_{n}$, not known in advance.
 In the PoissonL algorithm the mixed-membership vectors $\theta_n$ are known, i.e., for the $n$-$th$ gene we can write
\[
 \theta_{nk} = \left\{ \bv{rl} 1 & \hbox{if } k=j_n \\ 0 & \hbox{otherwise}, \ev \right.
\]
 where $j_n = \arg \min \bigm\{ L(n,k) : k \in [1,K] \bigm\}$.  This extension is similar in spirit to that introduced by Gaussian mixture to regular $K$-means.  In fact, we have
\[
 \theta_{nk} = Pr \bigm( cluster=k \bigm| data,parameters \bigm).
\]

 Note that introducing latent Dirichlet distributed mixed-membership vectors, $\theta_{n}$, ties together all the data in the inference task. This has the beneficial effect of reducing the variability of profile specific parameters as we make use of all the gene counts (independently of which profile they express the most) in estimating each such parameters.  Such an improvement in the estimates is expected \cite{Jame:Stei:1961}. 

 Our basic Poisson model is similar to that of \cite{Cann:2004}.  For a technical survey of related latent aspects models see \cite{Bunt:Jaku:2006}.

\section{Conclusions}
\label{sec:conclusions}

 In problems where features co-occur frequently (e.g., a gene can be present on multiple transcript, as picked up by SAGE), computational gains are hardly warranted.  Applications to problems that arise in computational biology, e.g., SAGE and microarray data, are one such case. 
 In this paper, we introduce probabilistic models to learn latent expression themes in an unsupervised fashion.  Our models capture the notion of ``contagion'' to characterize semantic themes underlying observed feature patterns, such as ``biological context'', within a hierarchical Bayesian scheme. We present model variants tailored to different properties of biological data, and we outline a general variational inference scheme for approximate posterior inference.

 Our results suggest the possibility of obtaining reasonable predictions of gene functions in an unsupervised fashion.  The estimates our models provide, in scenarios that feature realistic variability profiles for the data, are sharper than those entailed by existent methods based on stronger independence assumptions, and demonstrate feasibility of a promising hierarchical Bayesian formalism for soft clustering and latent aspect analysis.

\subsection*{Acknowledgments}

 The author's thinking about the issues discussed in this paper has benefited greatly from discussions and collaborations with William Cohen, at Carnegie Mellon University, and with David Blei, at Princeton University.
 This work was partially supported by National Institutes of Health (NIH) under Grant 1 R01 AG023141-01, by the Office of Naval Research (ONR) under Dynamic Network Analysis (N00014-02-1-0973), the National Science Foundation (NSF) and the Department of Defense (DOD) under MKIDS (IIS0218466).
 The views and conclusions contained in this document are those of the author and should not be interpreted as representing the official policies, either expressed or implied, of the NIH, the ONR, the NSF, the DOD, or the U.S. government.

\vskip 0.2in

\bibliographystyle{plainnat}

\begin{thebibliography}{21}
\providecommand{\natexlab}[1]{#1}
\providecommand{\url}[1]{\texttt{#1}}
\expandafter\ifx\csname urlstyle\endcsname\relax
  \providecommand{\doi}[1]{doi: #1}\else
  \providecommand{\doi}{doi: \begingroup \urlstyle{rm}\Url}\fi

\bibitem[Airoldi et~al.(2005)Airoldi, Cohen, and Fienberg]{Airo:Cohe:Fien:2005}
E.~M. Airoldi, W.~W. Cohen, and S.~E. Fienberg.
\newblock Bayesian models for frequent terms in text.
\newblock In \emph{Proceedings of the Classification Society of North America
  and INTERFACE Annual Meetings}, 2005.

\bibitem[Airoldi et~al.(2006{\natexlab{a}})Airoldi, Anderson, Fienberg, and
  Skinner]{Airo:Ande:Fien:Skin:2006}
E.~M. Airoldi, A.~G. Anderson, S.~E. Fienberg, and K.~K. Skinner.
\newblock Who wrote {R}onald {R}eagan's radio addresses?
\newblock \emph{Bayesian Analysis}, 1\penalty0 (2):\penalty0 289--320,
  2006{\natexlab{a}}.

\bibitem[Airoldi et~al.(2006{\natexlab{b}})Airoldi, Fienberg, Joutard, and
  Love]{Airo:Fien:Jout:Love:2006}
E.~M. Airoldi, S.~E. Fienberg, C.~Joutard, and T.~M. Love.
\newblock Discovering latent patterns with hierarchical {B}ayesian
  mixed-membership models and the issue of model choice.
\newblock Technical Report CMU-ML-06-101, School of Computer Science, Carnegie
  Mellon University, April 2006{\natexlab{b}}.

\bibitem[Blackshaw et~al.(2004)Blackshaw, Harpavat, Trimarchi, Cai, Huang, Kuo,
  Fraioli, Cho, Yung, and Asch]{Blac:Harp:Trim:Cai:2004}
S.~Blackshaw, S.~Harpavat, J.~Trimarchi, L.~Cai, H~Huang, W.~P. Kuo, R.~E.
  Fraioli, S.~H. Cho, R.~Yung, and E.~Asch.
\newblock Genomic analysis of mouse retinal development.
\newblock \emph{PLoS Biology}, 2004.

\bibitem[Blei et~al.(2003)Blei, Ng, and Jordan]{Blei:Ng:Jord:2003}
D.~M. Blei, A.~Ng, and M.~I. Jordan.
\newblock Latent {D}irichlet allocation.
\newblock \emph{Journal of Machine Learning Research}, 3:\penalty0 993--1022,
  2003.

\bibitem[Buntine and Jakulin(2004)]{Bunt:Jaku:2004}
W.~Buntine and A.~Jakulin.
\newblock Applying discrete {PCA} in data analysis.
\newblock In \emph{Uncertainty in Artificial Intelligence}, 2004.

\bibitem[Buntine and Jakulin(2006)]{Bunt:Jaku:2006}
W.~L. Buntine and A.~Jakulin.
\newblock Discrete components analysis.
\newblock In C.~Saunders, M.~Grobelnik, S.~Gunn, and J.~Shawe-Taylor, editors,
  \emph{Subspace, Latent Structure and Feature Selection Techniques}.
  Springer-Verlag, 2006.
\newblock to appear.

\bibitem[Cai et~al.(2004)Cai, Huang, Blackshaw, Liu, Cepko, and
  Wong]{Cai:Huan:Blac:Liu:2004}
L.~Cai, H.~Huang, S.~Blackshaw, J.~S. Liu, C.~L. Cepko, and W.~H. Wong.
\newblock Clustering analysis of {SAGE} data using a {Poisson} approach.
\newblock \emph{Genome Biology}, 5\penalty0 (7):\penalty0 R51, 2004.

\bibitem[Canny(2004)]{Cann:2004}
J.~Canny.
\newblock {G}a{P}: A factor model for discrete data.
\newblock In \emph{Proceedings of the 27th Annual International ACM SIGIR
  Conference on Research and Development in Information Retrieval}, 2004.

\bibitem[Carlin and Louis(2005)]{Carl:Loui:2005}
B.~P. Carlin and T.~A. Louis.
\newblock \emph{Bayes and {E}mpirical {B}ayes Methods for Data Analysis}.
\newblock Chapman \& Hall, second edition, 2005.

\bibitem[Cohn and Hofmann(2001)]{Cohn:Hofm:2001}
D.~Cohn and T.~Hofmann.
\newblock The missing link---{A} probabilistic model of document content and
  hypertext connectivity.
\newblock In \emph{Advances in Neural Information Processing Systems 13}, 2001.

\bibitem[Griffiths and Steyvers(2004)]{Grif:Stey:2004}
T.~L. Griffiths and M.~Steyvers.
\newblock Finding scientific topics.
\newblock \emph{Proceedings of the National Academy of Sciences}, 101\penalty0
  (Suppl. 1):\penalty0 5228--5235, 2004.

\bibitem[James and Stein(1961)]{Jame:Stei:1961}
W.~James and C.~M. Stein.
\newblock Estimation with quadratic loss.
\newblock In \emph{Proceedings of the 4th Berkeley Symposium on Mathematical
  Statistics and Probability}, volume~1, pages 361--379, 1961.

\bibitem[Johnson et~al.(1992)Johnson, Kotz, and Kemp]{John:Kotz:Kemp:1992}
N.~L. Johnson, S.~Kotz, and A.~W. Kemp.
\newblock \emph{Univariate Discrete Distributions}.
\newblock John Wiley, 1992.

\bibitem[Kadane et~al.(2006)Kadane, Shmueli, Minka, Borle, and
  Boatwright]{Kada:Shmu:Mink:etal:2006}
J.~B. Kadane, G.~Shmueli, T.~P. Minka, S.~Borle, and P.~Boatwright.
\newblock Conjugate analysis of the {C}onway-{M}axwell-{P}oisson distribution.
\newblock \emph{Bayesian Analysis}, 1\penalty0 (2):\penalty0 363--374, 2006.

\bibitem[Minka and Lafferty(2002)]{Mink:Laff:2002}
T.~Minka and J.~Lafferty.
\newblock Expectation-propagation for the generative aspect model.
\newblock In \emph{Uncertainty in Artificial Intelligence}, 2002.

\bibitem[Pritchard et~al.(2000)Pritchard, Stephens, and
  Donnelly]{Prit:Step:Donn:2000}
J.~Pritchard, M.~Stephens, and P.~Donnelly.
\newblock Inference of population structure using multilocus genotype data.
\newblock \emph{Genetics}, 155:\penalty0 945--959, 2000.

\bibitem[Rosenberg et~al.(2002)Rosenberg, Pritchard, Weber, Cann, Kidd,
  Zhivotovsky, and Feldman]{Rose:Prit:Webe:etal:2002}
N.~A. Rosenberg, J.~K. Pritchard, J.~L. Weber, H.~M. Cann, K.~K. Kidd, L.~A.
  Zhivotovsky, and M.~W. Feldman.
\newblock Genetic structure of human populations.
\newblock \emph{Science}, 298:\penalty0 2381--2385, 2002.

\bibitem[Simon(1955)]{Simo:1955}
H.~A. Simon.
\newblock On a class of skew distribution functions.
\newblock \emph{Biometrika}, 42:\penalty0 425--440, 1955.

\bibitem[Vesculescu et~al.(1995)Vesculescu, Zhang, Vogelstein, and
  Kinzler]{Vesc:Zhan:Voge:Kinz:1995}
V.~E. Vesculescu, L.~Zhang, B.~Vogelstein, and K.~W. Kinzler.
\newblock Serial analysis of gene expression.
\newblock \emph{Science}, 270:\penalty0 484--487, 1995.

\bibitem[Xing et~al.(2003)Xing, Jordan, Karp, and
  Russell]{Xing:Jord:Karp:Russ:2003}
E.~P. Xing, M.~I. Jordan, R.~M. Karp, and S.~Russell.
\newblock A hierarchical {B}ayesian markovian model for motifs in biopolymer
  sequences.
\newblock In \emph{Advances in Neural Information Processing Systems},
  volume~16, 2003.

\end{thebibliography}

\end{document}